\documentclass{elsart}

\usepackage{graphicx}
\usepackage{rotate}

\usepackage{amssymb}

\begin{document}

\begin{frontmatter}



\title{Synchronization processes in complex networks}


\author[urv]{Alex Arenas}, \author[ub]{Albert D{\'\i}az-Guilera} and \author[ub]{Conrad J. P{\'e}rez-Vicente}

\address[urv]{Departament d'Enginyeria Inform{\`a}tica i Matem{\`a}tiques,
  Universitat Rovira i Virgili, 43007 Tarragona, Spain}

\address[ub]{Departament de F{\'\i}sica Fonamental, Universitat de
  Barcelona, Mart{\'\i} i Franqu{\`e}s 1, 08028 Barcelona, Spain}

\begin{abstract}
We present an extended analysis, based on the dynamics towards synchronization
of a system of coupled oscillators, of the hierarchy of communities in complex networks. In the synchronization process, different structures corresponding to well defined communities of nodes appear in a hierarchical way. The analysis also provides a useful connection between synchronization dynamics, complex networks topology and spectral graph analysis.
\end{abstract}

\end{frontmatter}


\section{Introduction}

In 1998 Watts and Strogatz presented a simple model of network's
structure that was the seed of the modern theory of complex
networks \cite{watts}. Beginning with a regular lattice, they
showed that the addition of a small number of random links reduces
the diameter drastically. This effect, know as small-world
effect, was detected in natural and artificial networks. The
research was in part originally inspired by Watts' efforts to understand
the synchronization of cricket chirps, which show a high degree of
coordination over long distances as though the insects where
``invisibly" connected. Since then complex networks are being 
subject of attention of the physicists' community
\cite{strogatz,barev,newmanrev,yamirrep}.

Complex networks are found in fields as diverse as the Internet,
the World-Wide-Web, food-webs, and many forms of biological and
social organizations (see \cite{buchanan} and references therein).
The description of these networks, as it occurs in many physical
systems, can be performed at different scales. At the lower level
of description, the ``microscale"  is
represented by single nodes. 
From the static point of view the key point is to determine
certain properties of individual
nodes (degree, centrality, clustering, etc.), while
from the dynamics of the network point of view, the important issue is to know about the
dynamical process each node is performing. This level of
description is unfortunately very precise and does not allow a
generic analysis of the global properties of the system. At the
other extreme, we have the higher level of description, the
``macroscale", represented by the statistical
properties of the network as a whole. This description has been
the realm of statistical physics in complex networks and has
provided great insight in the universality of certain
features of many real world systems.

In the middle of these descriptions still remains a huge space for
different scales of descriptions that we like to name as
``mesoscales", or intermediate scales. These scales are understood
as substructures (eventually subgraphs) that have topological
entity compared to the whole network, e.g. motifs
\cite{alon1,alon2}, cliques \cite{palla}, cores \cite{almaas},
loops \cite{bianconi} or, generally speaking, communities
\cite{newmanepjb}. In particular, the community detection problem
concerning the determination of mesoscopic structures that have
functional, relational or even social entity is still
controversial, starting  from the ``a priori" definition of what a
community is \cite{faust}. The correct determination of the
mesoscale in complex networks is a major challenge. Under the name
of the community detection problem, consisting in finding a 'good'
partition of the network in sub-graphs that represent communities
according to a given definition, physicists have provided
different methods that confront this challenge\cite{jstat}.
However, in many complex networks the organization of nodes is not
completely represented by a unique partition but by a set of
nested communities that appear at different topological scales.
This evidence comes from indirect experimental data revealing
functionalities in complex networks that involve different subsets
of nodes at different hierarchical levels \cite{raso,rives}.

In a completely different scenario, physicists have largely
studied the dynamics of complex biological systems, and in
particular the paradigmatic analysis of large populations of
coupled oscillators \cite{winfree,strogatzsync,kurabook}. The
connection between the study of synchronization processes and
complex networks is interesting by itself. Indeed, the original
inspiration of Watts and Strogatz in the development of the
Small-World network structure was, as mentioned before,  to
understand the synchronization of cricket chirps. This
synchronization phenomena as many others e.g. asian fireflies
flashing at unison, pacemaker cells in the heart oscillating in
harmony, etc. have been mainly described under the mean field
hypothesis that assumes that all oscillators behave identically
and interact with the rest of the population. Recently, the
emergence of synchronization phenomena in complex networks has been
shown to be closely related to the underlying topology of
interactions  \cite{atay} beyond the macroscopic description.

In this paper we study of the dynamics towards synchronization in
complex networks at the mesoscale description, extending previous results \cite{nos}. The paper is
structured as follows: in section II we present the
synchronization model studied. In section III we describe a method
to construct synthetic networks with a well prescribed
hierarchical community structure. In section IV, we expose the
analysis of the route towards synchronization and their
relationship with the topological structure. Finally, we conclude with a discussion about the
synchronization processes in complex networks.

\section{Synchronization: Kuramoto's model}

One of the most successful attempts to understand synchronization
phenomena was due to Kuramoto \cite{kurabook}, who analyzed a
model of phase oscillators coupled through the sine of their phase
differences. The model is rich enough to display a large variety
of synchronization patterns and sufficiently flexible to be
adapted to many different contexts \cite{conradrev}. The Kuramoto
model consists of a population of $N$ coupled phase oscillators
where the phase of the $i$-th unit, denoted by $\theta_i(t)$,
evolves in time according to the following dynamics
\begin{equation}
\frac{d\theta_i}{dt}=\omega_i + \sum_{j}
K_{ij}\sin(\theta_j-\theta_i) \hspace{0.5cm} i=1,...,N
 \label{ks}
\end{equation}
\noindent where $\omega_i$ stands for its natural frequency and
$K_{ij}$ describes the coupling between units. The original model
studied by Kuramoto assumed mean-field interactions $K_{ij}=K,
\forall i,j$. In absence of noise the long time properties of the
population are determined by analyzing the only two factors which
play a role in the dynamics: the strength of the coupling $K$
whose effect tends to synchronize the oscillators (same phase)
versus the width of the distribution of natural frequencies, the
source of disorder which drives them to stay away each other by
running at different velocities. For unimodal distributions, there
is a critical coupling $K_c$ above which synchronization emerges
spontaneously.

Recently, due to the realization that many networks in nature have
complex topologies, these studies have been extended to systems where the 
pattern of connections is local but not necessarily regular
\cite{barahona,motter1,yamir,hong,motter2,lee,munozprl,chavez,restrepo1,restrepo2}.
Usually, due to the complexity of the analysis some further
assumptions have been introduced. For instance, it has been a
normal practice to assume that the oscillators are identical.
Obviously, in absence of disorder, i.e. if  $(\omega_i = \omega
~\forall i)$ there is only one attractor of the dynamics: the
fully synchronized regime where $\theta_i = \theta, ~\forall i$.
In this context the interest concerns not the final locked state
in itself but the route to the attractor. In particular, it has
been shown \cite{yamir2,kahng} that high densely interconnected
sets of oscillators (motifs) synchronize more easily that those
with sparse connections. This scenario suggests that for a complex
network with a non-trivial connectivity pattern, starting from
random initial conditions, those highly interconnected units
forming local clusters will synchronize first and then, in a
sequential process, larger and larger spatial structures also will
do it up to the final state where the whole population should have
the same phase. We expect this process to occur at different time
scales if a clear community structure exists.  Thus, the dynamical
route towards the global attractor will reveal different
topological structures, presumably those which represent
communities. Therefore, it is the complete dynamical process what
unveils the whole organization at all scales, from the microscale
at a very early stages up to the macroscale at the end of the time
evolution. On the contrary, those systems endowed with a regular
topological structure will usually display a trivial dynamics with a
single time scale for synchronization, although some recent studies indicate also other possibilities \cite{wiley}.

It is a normal practice to define, for the Kuramoto model, a
global order parameter to characterize the level of entrainment
between oscillators. The normal choice is to use the following
complex-valued order-parameter
\begin{equation}
 r e^{i\psi}=\frac{1}{N}\sum_{j=1}^{N} e^{i\theta_{j}}.
\end{equation}
where $r(t)$ with $0\leq r(t)\leq 1$ measures the coherence of the
oscillator population, and $\psi(t)$ is the average phase.
However, this definition, although suitable for mean-field models
is not efficient to identify local dynamic effects. In particular
it does not give information about the route to the attractor
(fully synchronization) in terms of local clusters which is so
important to identify functional groups or communities. For this
reason, instead of considering a global observable, we define a
local order parameter measuring the average of the correlation
between pairs of oscillators
 \begin{equation}
\rho_{ij}(t)=<cos(\theta_i(t)-\theta_j(t))>
 \label{ro}
\end{equation}
\noindent where the brackets stand for the average over initial random phases.
The main advantage of this approach is that it allows to trace the time evolution of pairs of
oscillators and therefore to identify compact clusters reminiscent
of the existence of communities.

\section{Structured complex networks}

To give evidence of the aforementioned facts we have analyzed the
dynamics towards synchronization --time evolution of
$\rho_{ij}(t)$-- in computer-generated graphs with community structure.

The paradigmatic model of network with a well defined community structure that
has been used as a benchmark for different community detection
algorithms \cite{jstat}, was proposed by Girvan and Newman
\cite{girvannewman}. In that model the authors construct a network
of 128 nodes as a set of 4 communities, each one formed by  32
nodes. Fixing the mean number of links per node at a value of 16,
the parameter describing the sharpness of the community
distribution is $z_{in}$, the average number of links within the
community. Here, we propose two generalizations of this model:

\begin{itemize}
\item Including communities of different sizes. In
\cite{leon} it is shown that the algorithms for detecting the
community structure are very sensitive to the size of the communities
themselves, and a model to construct networks with inhomogeneous distributions of communities
is proposed. In this case the networks are
parametrized by two quantities, the internal and the external
cohesion. As an example of such network see Fig.\ref{inhomo}a,
with a clear distinction between the communities of different
sizes.


\begin{figure}
\begin{tabular}{|c|c|}
\hline\\
a)\includegraphics*[width=0.44\columnwidth]{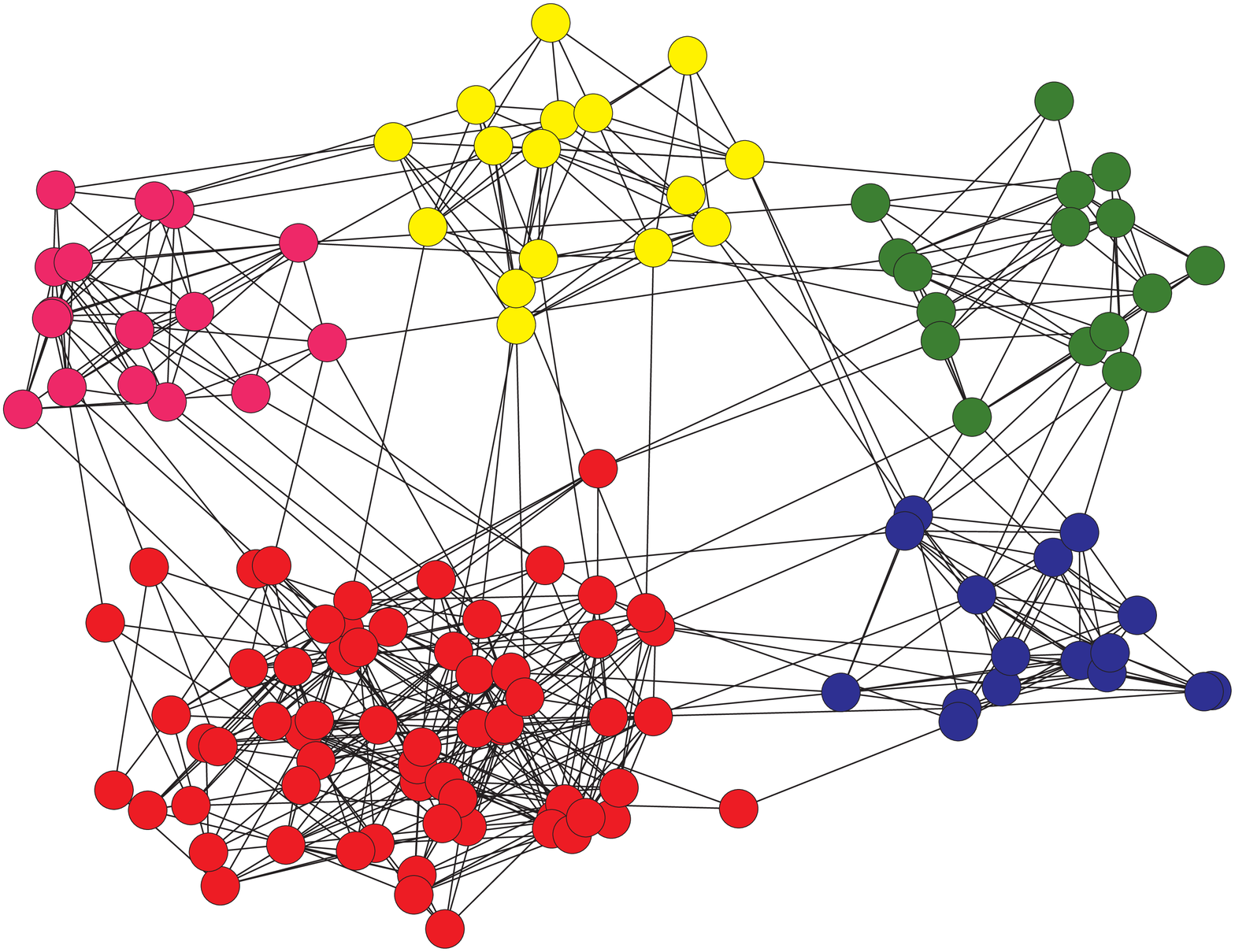} &
b)\includegraphics*[width=0.44\columnwidth]{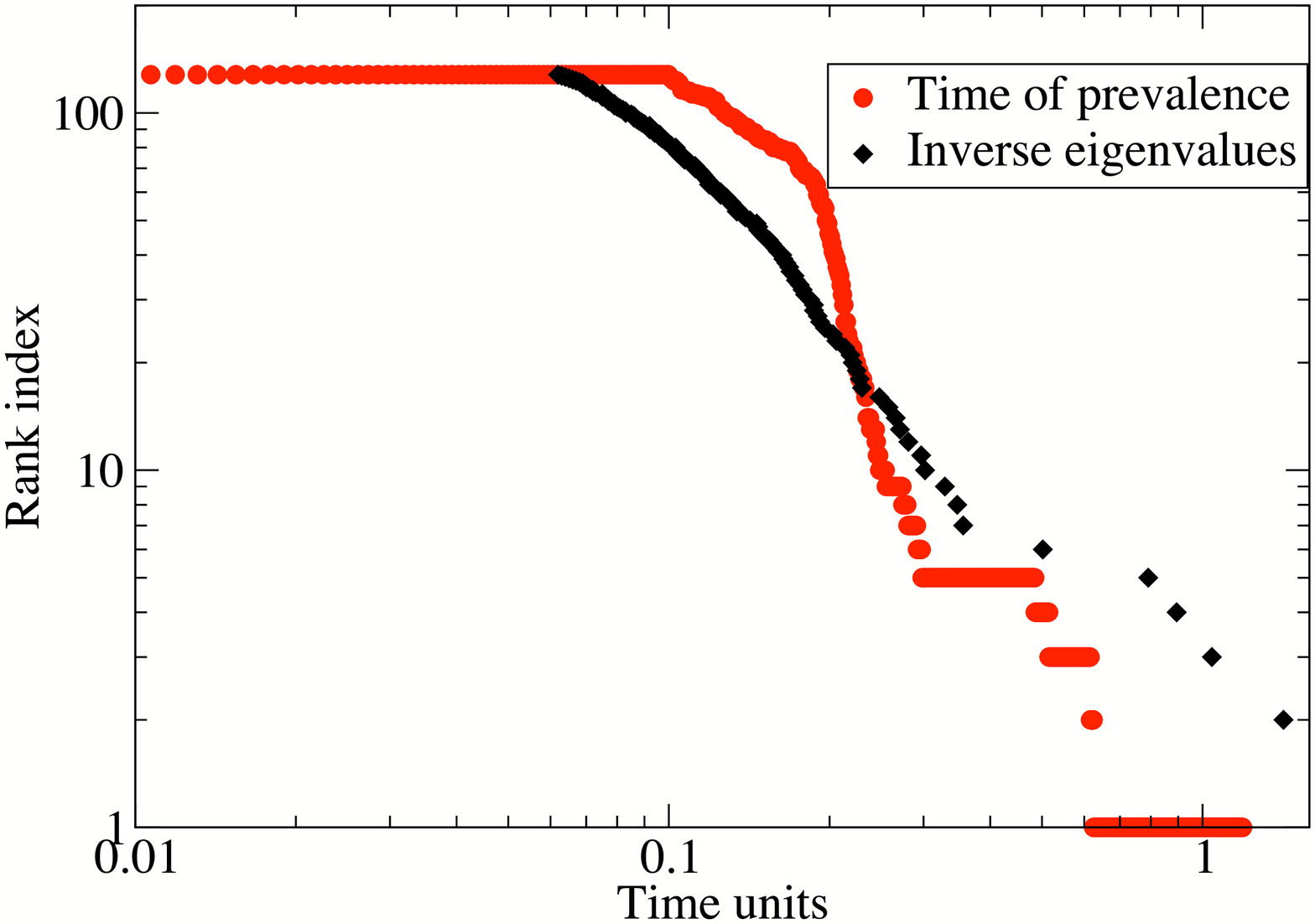} \\ \hline
c)\includegraphics*[width=0.44\columnwidth]{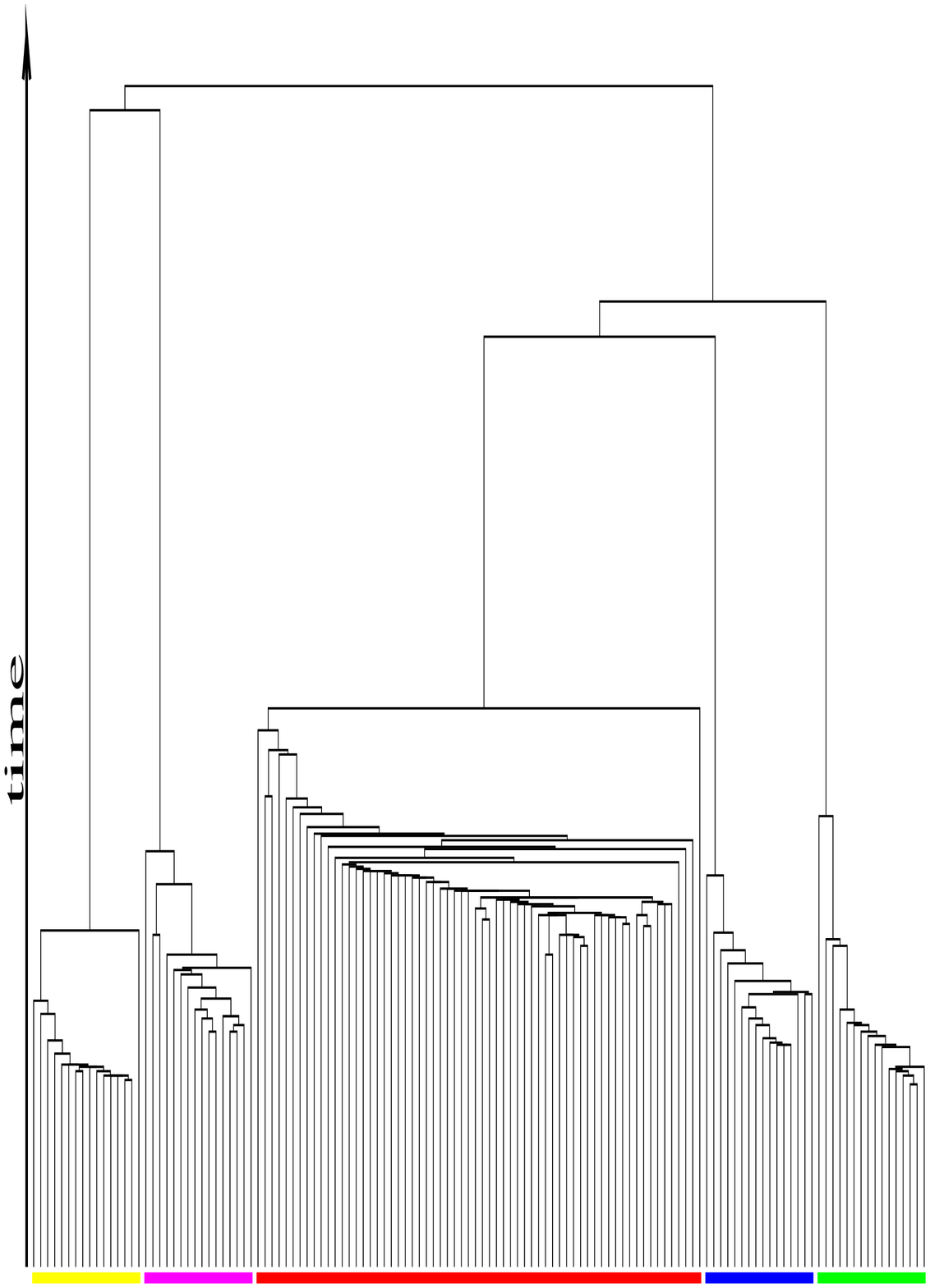} &
d)\includegraphics*[width=0.44\columnwidth]{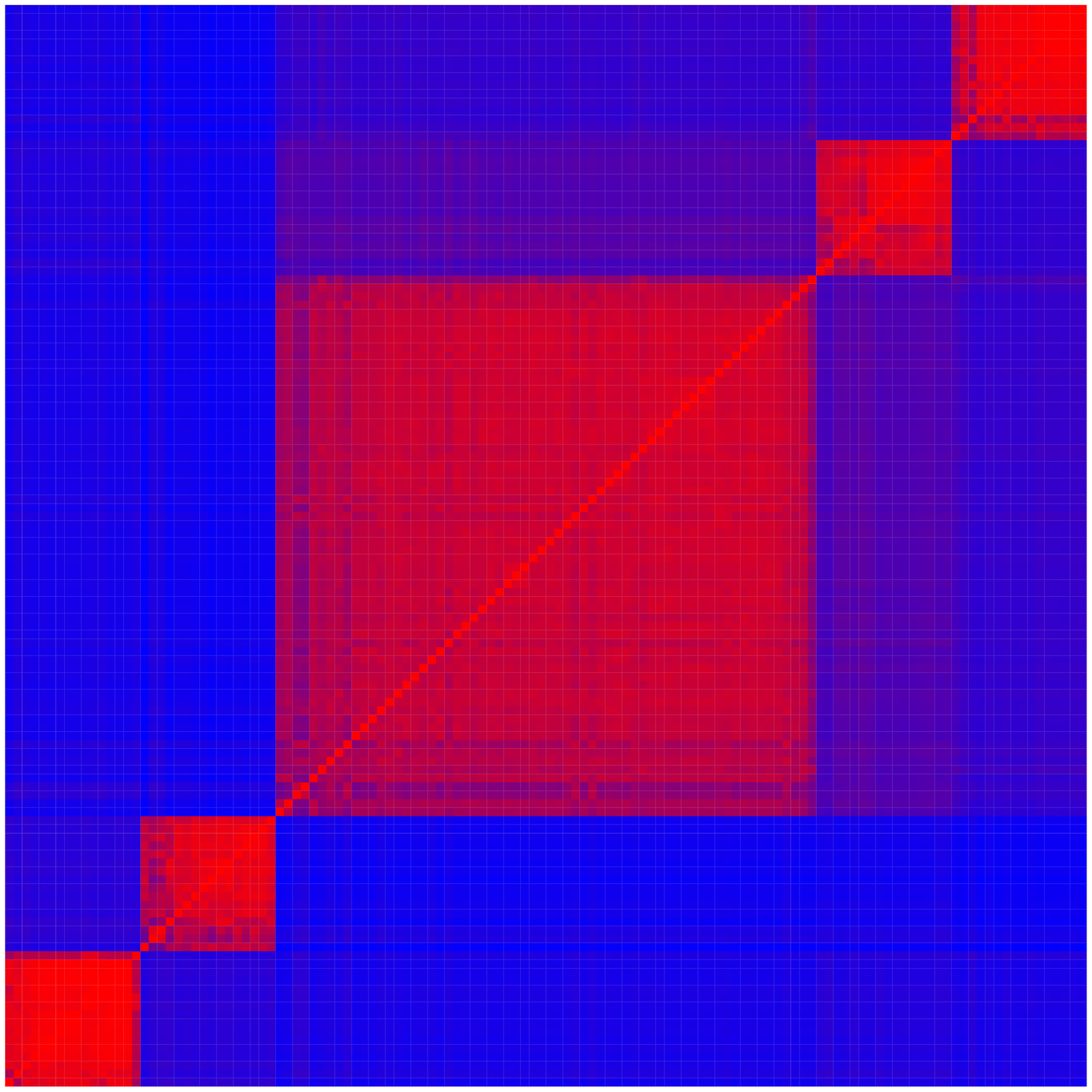}\\
\hline
\end{tabular}
 \caption{ Network with a inhomogeneous distribution of communities. a) the network structure; b) eigenvalues spectra and number of detected communities as a function of time; c) dendogram of the community merging. d) time needed for each pair of oscillators to synchronize. Red for shorter times, blue for larger times.}
\label{inhomo}
\end{figure}

\item Including several
hierarchical levels of communities. We take a set of $N$ nodes
and divide it into $n_1$ groups of equal size; each of these
groups is then divided into $n_2$ groups and so on up to a number
of steps $k$ which defines the number of hierarchical levels of
the network. Then we add links to the networks in such a way that
at each node we assign at random a number of $z_{1}$ neighbours
within its group at the first level, $z_{2}$ neighbours within
the group at the second level and so on. There is a remaining
numbers of links that each node has to the rest of the network,
that we will call $z_{out}$. In this case it is easy to compute
the modularity of the partition \cite{girvannewman} at any level $l\le k$
\begin{equation}
Q_{n_1\cdot n_2 \cdot \ldots \cdot n_l}=\frac{z_l+\ldots  + z_k}{z_{out}+z_1+\ldots + z_k}-\frac{1}{n_1 \cdot n_2 \ldots \cdot n_l}
\end{equation}
and its numerical value tells us how good
as partition into a given community structure is. 
In \cite{nos} we considered networks with 
two hierarchical levels with 256 nodes, and $n_1=n_2=4$;
this gives two possible partitions: one with 4 communities and the
other one with 16 communities. In that case one can relate the more stable regions
with larger values of modularity.
Here we consider a network with 3 levels with 64 nodes for which $n_1=n_2=n_3=2$, see Fig.\ref{3n}a.
\end{itemize}


\begin{figure}
\begin{tabular}{|c|c|}
\hline\\
a)\includegraphics*[width=0.44\columnwidth]{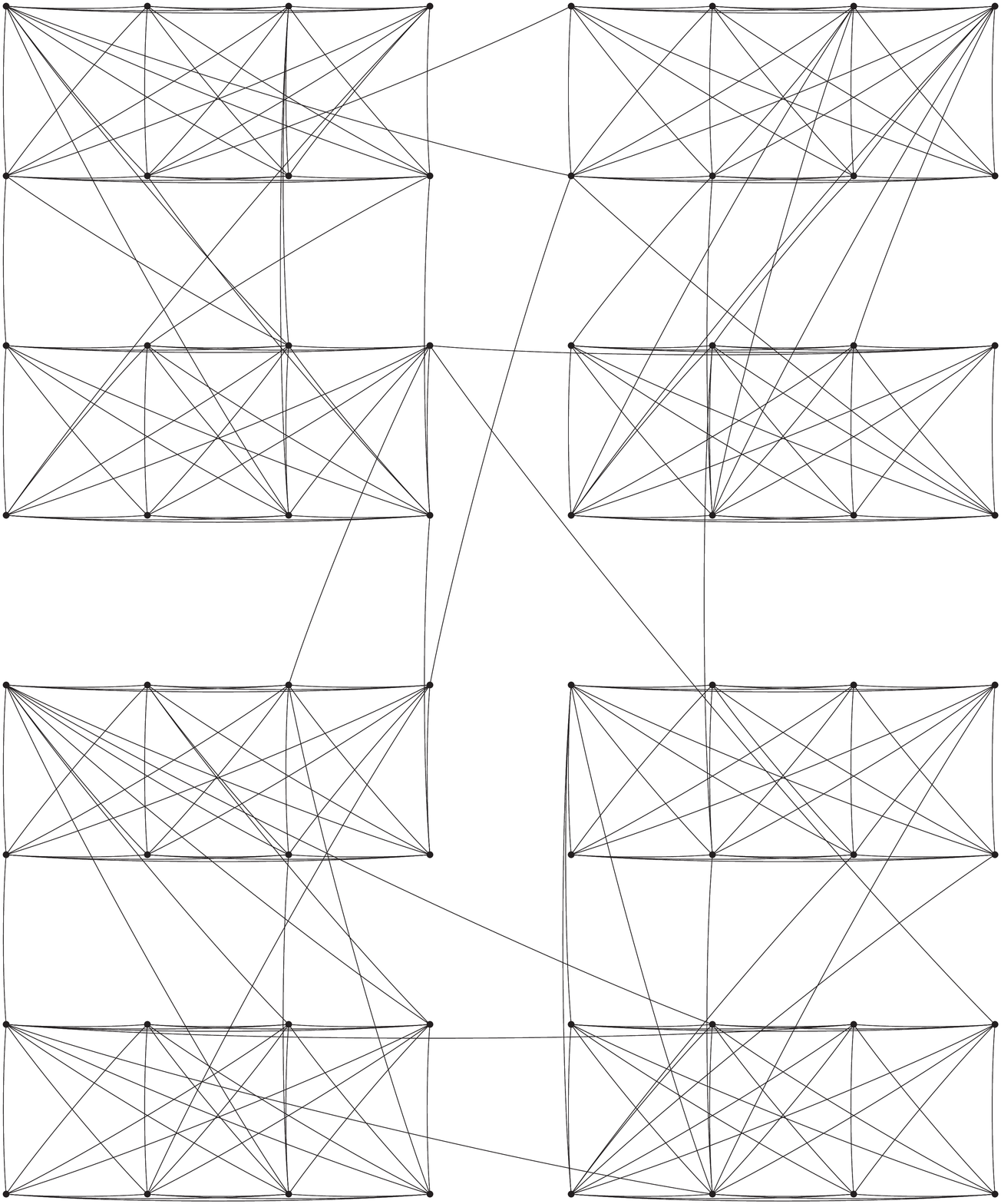} &
b)\includegraphics*[width=0.44\columnwidth,angle=-0]{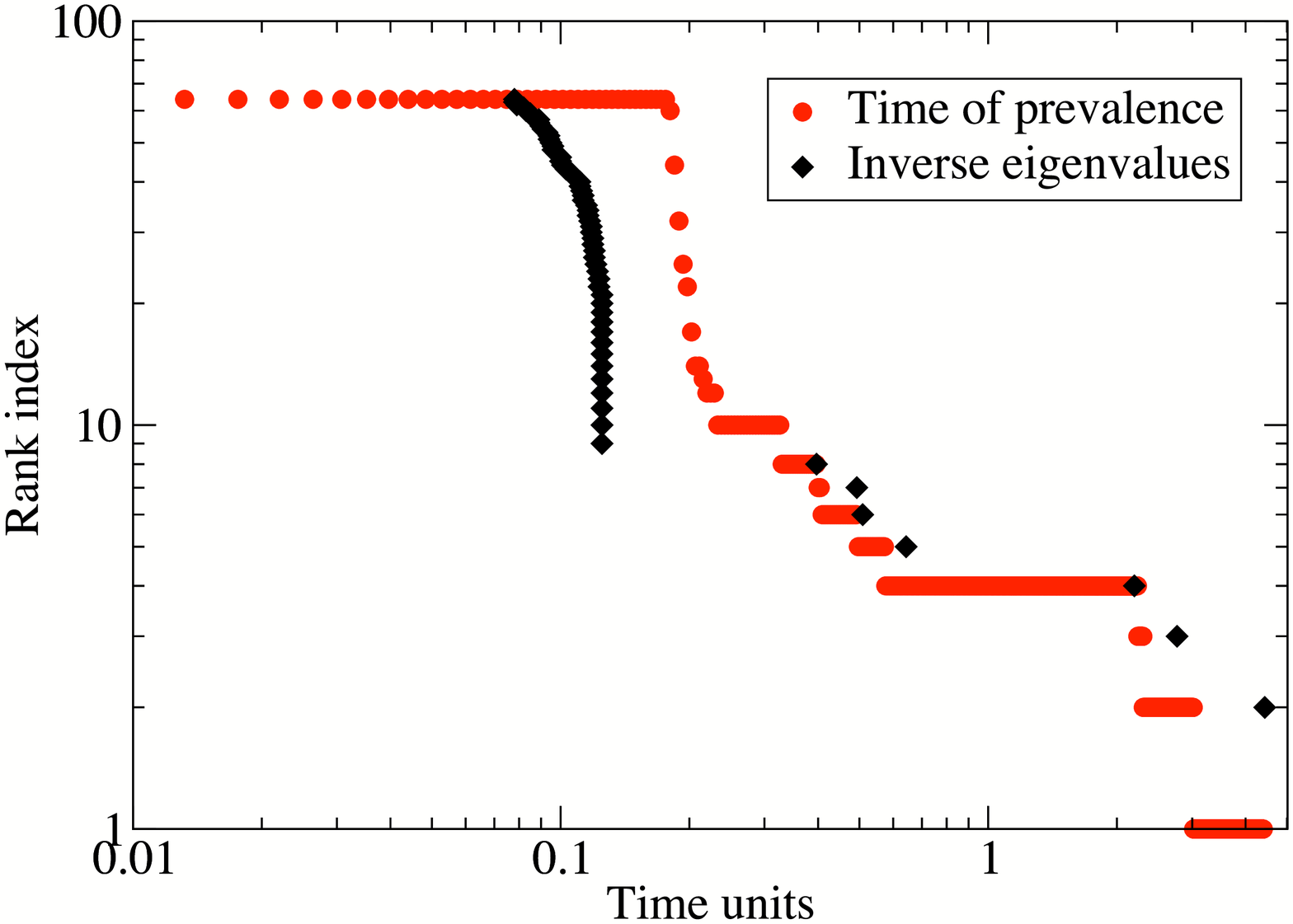} \\ \hline
c)\includegraphics*[width=0.44\columnwidth]{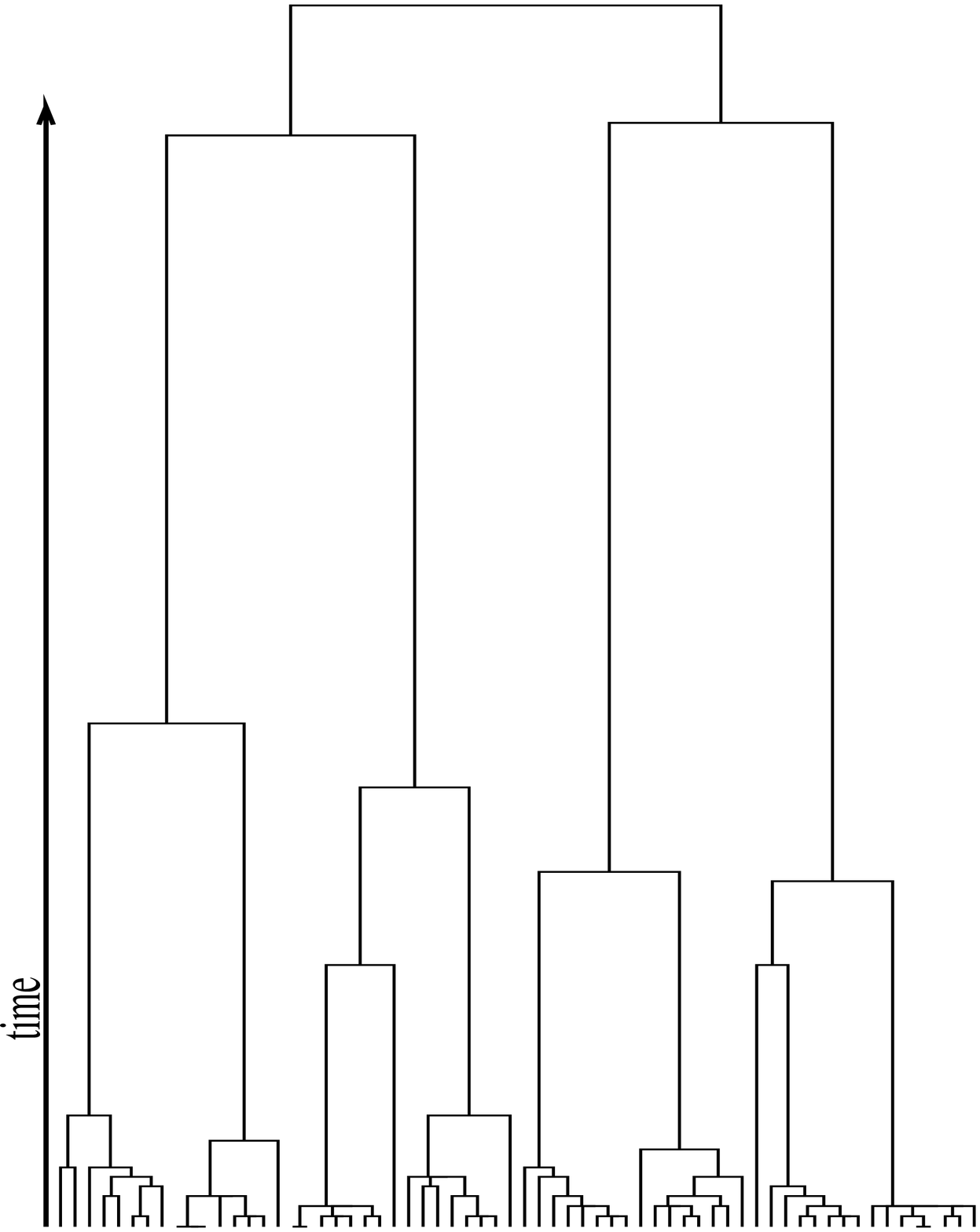} &
d)\includegraphics*[width=0.44\columnwidth]{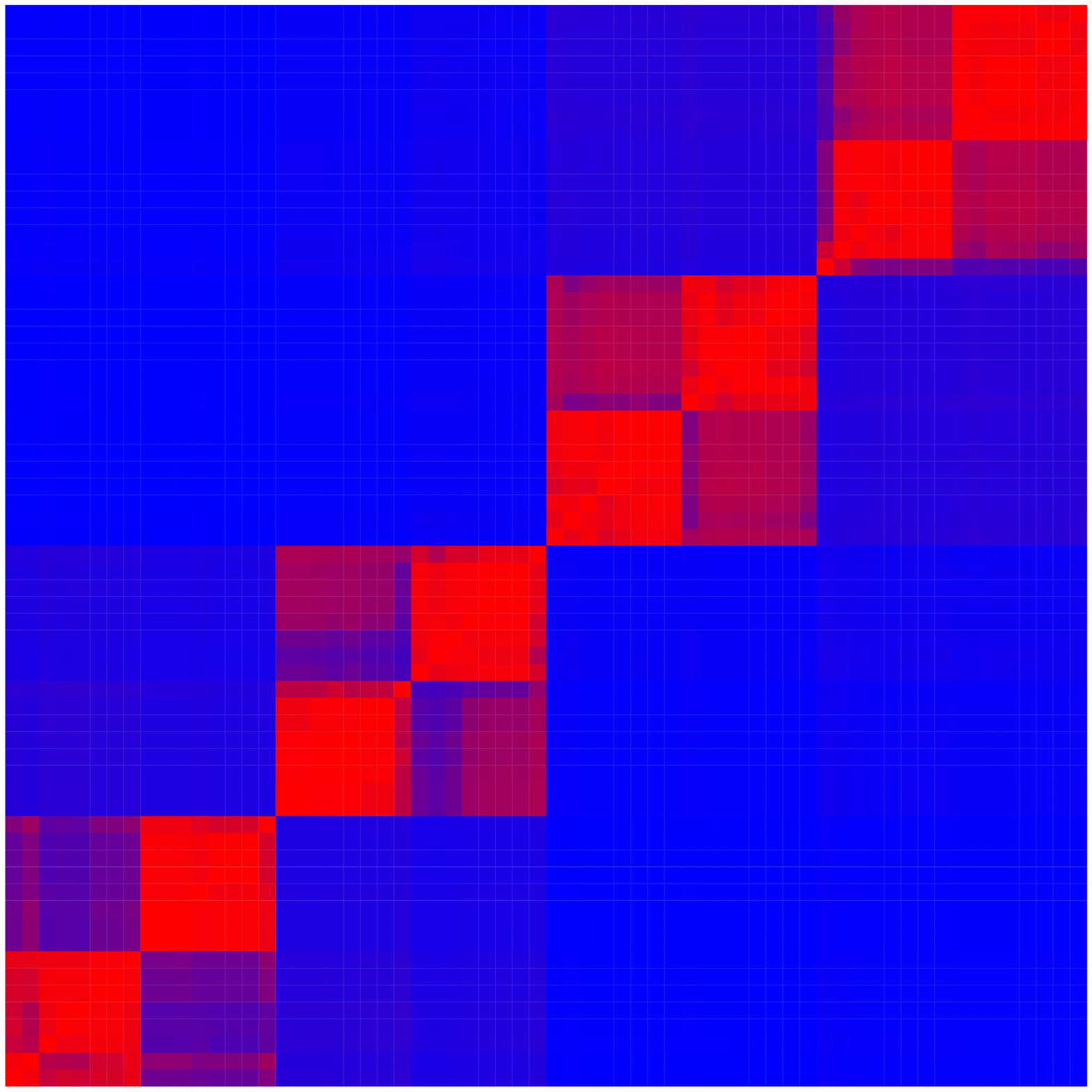} \\
\hline
\end{tabular}
 \caption{ Network with 3 levels of communities. a) the network structure; b)eigenvalues spectra and number of detected communities as a function of time; c) dendogram of the community merging; d) time needed for each pair of oscillators to synchronize. Red for shorter times, blue for larger times.}
\label{3n}
\end{figure}

\begin{figure}
\begin{tabular}{|c|c|}
\hline\\
a)\includegraphics*[width=0.47\columnwidth]{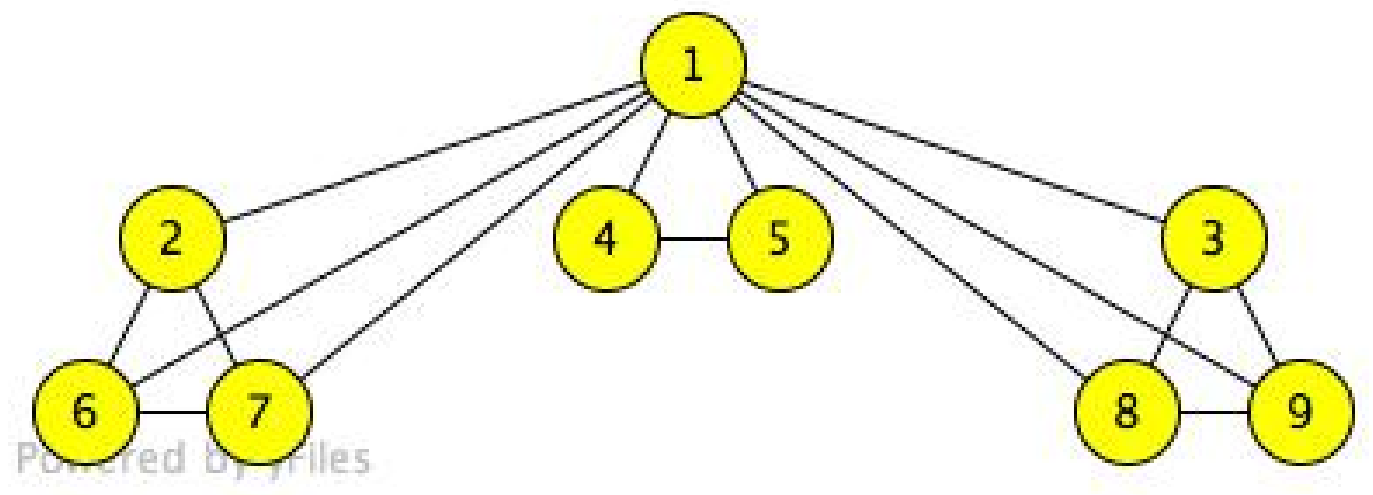} &
b)\includegraphics*[width=0.47\columnwidth]{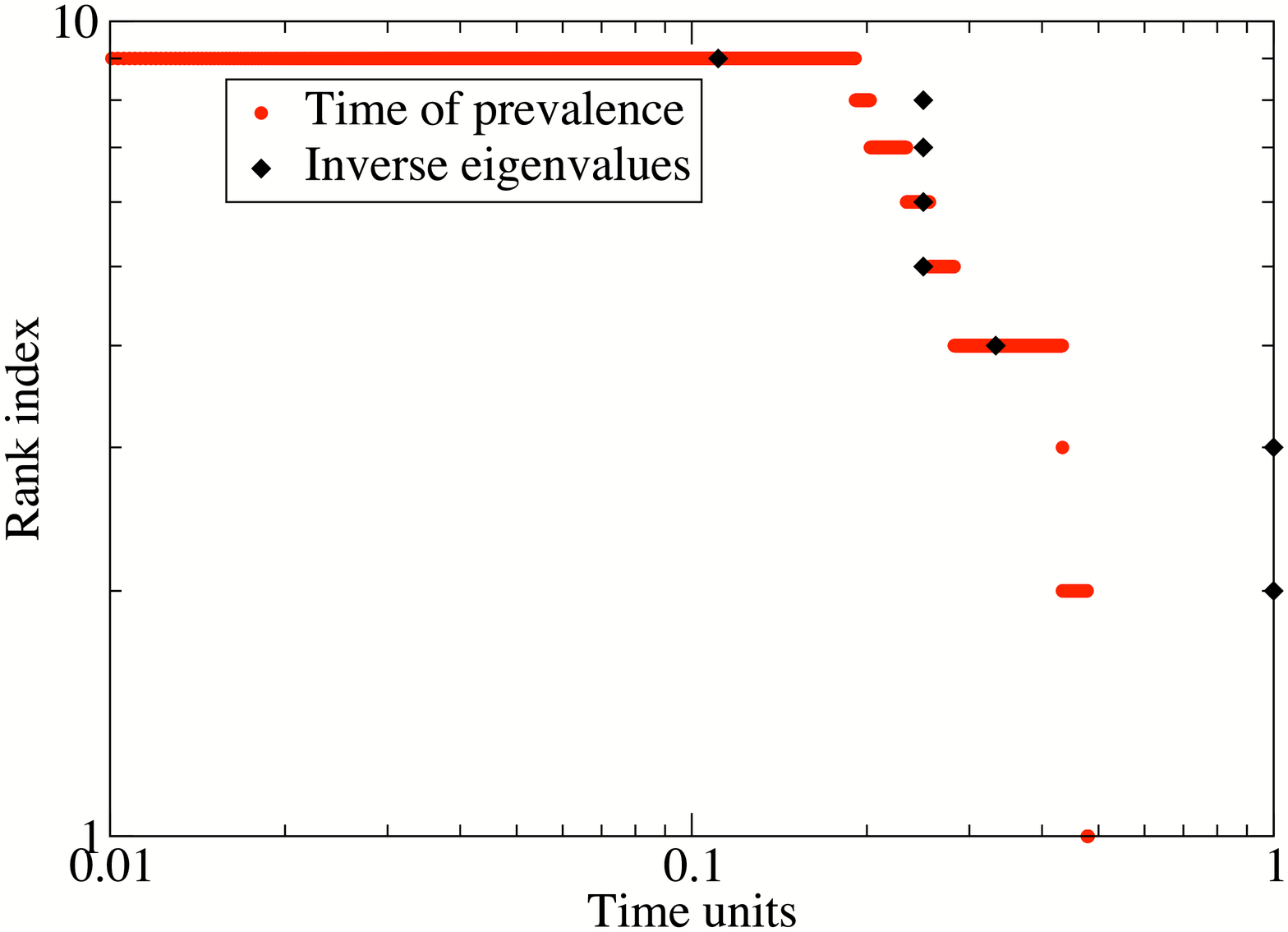} \\ \hline
c)\includegraphics*[width=0.47\columnwidth]{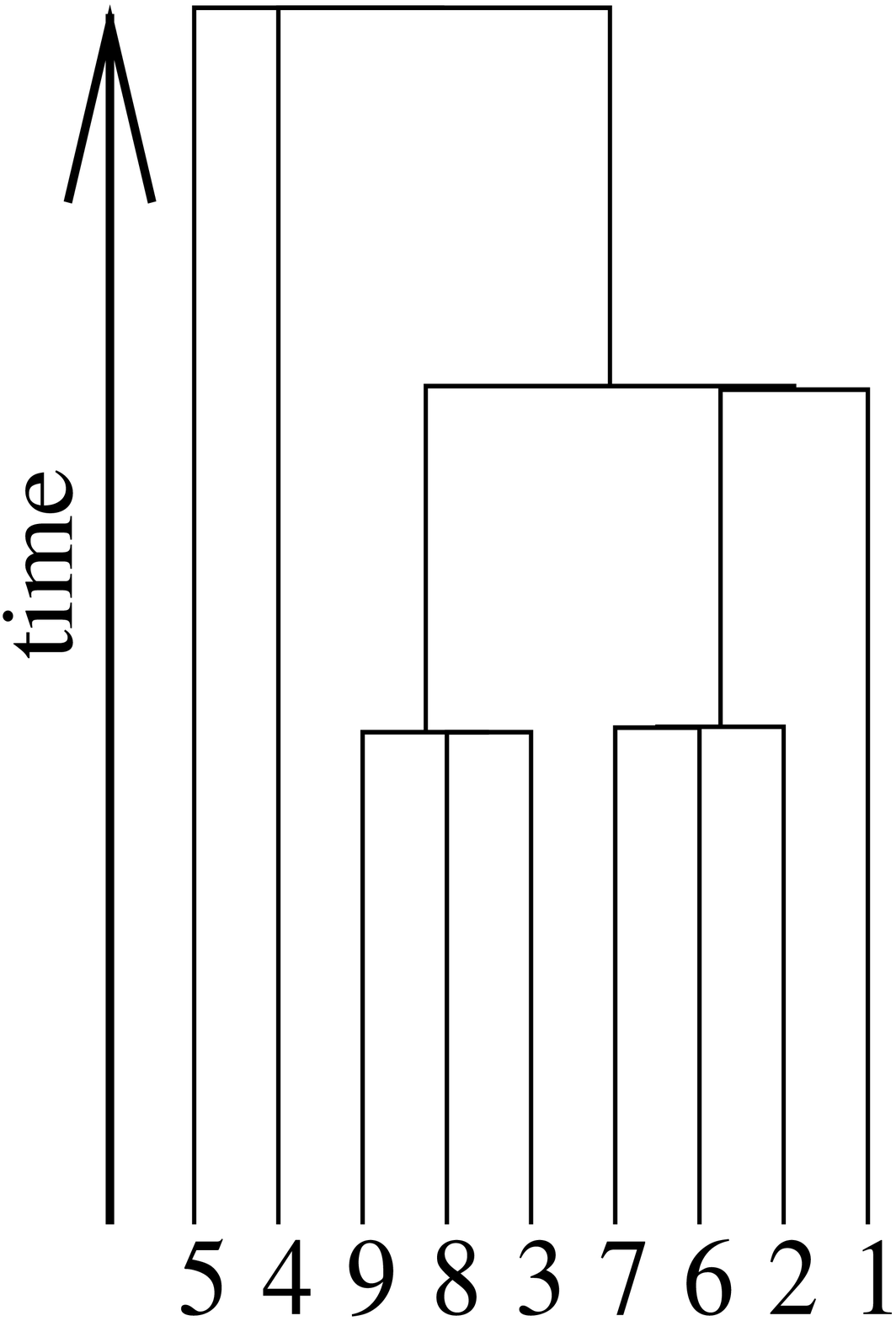} &
d)\includegraphics*[width=0.47\columnwidth]{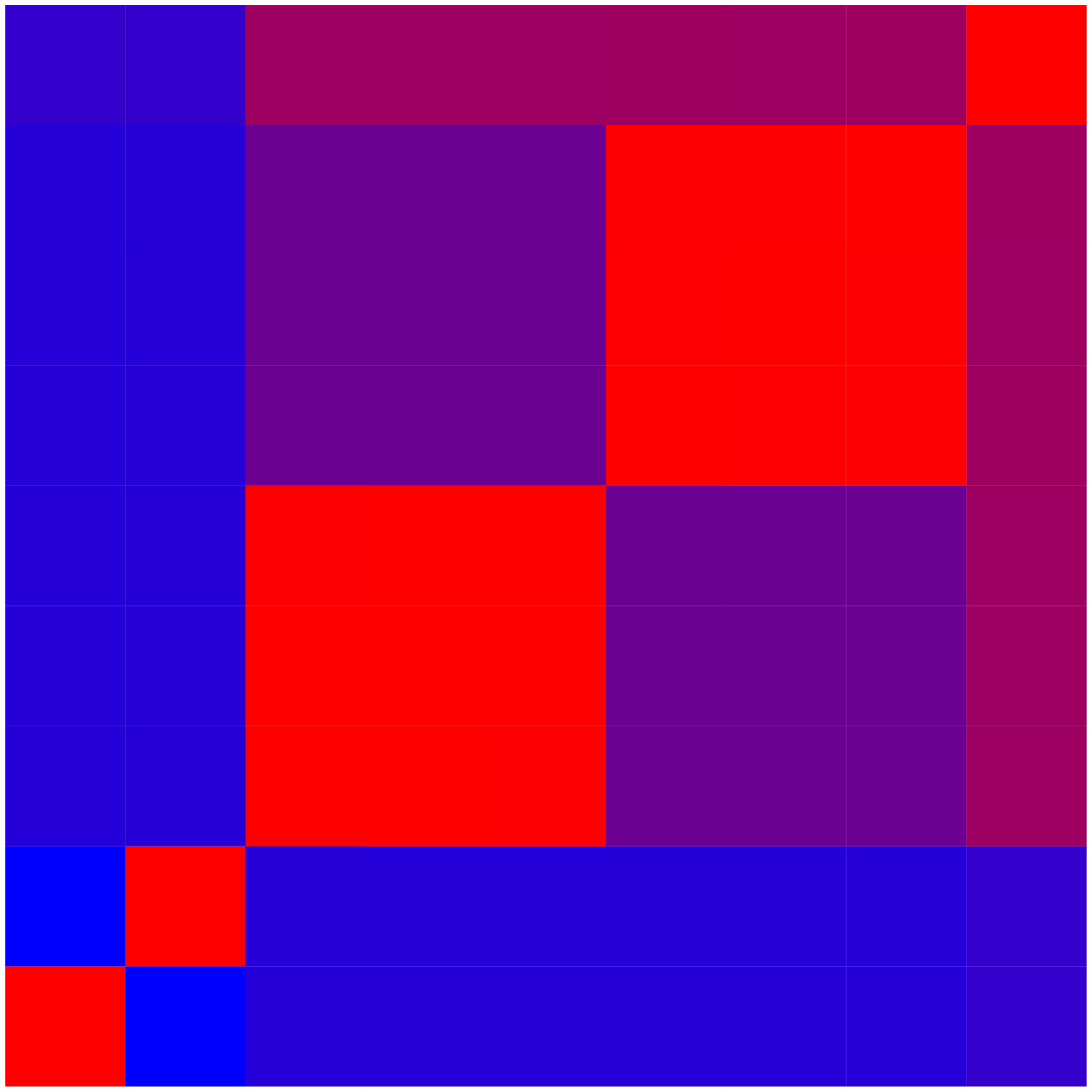}\\
\hline
\end{tabular}
  \caption{ Deterministic network with 2 levels. a) the network structure; b) eigenvalues spectra and number of detected communities as a function of time; c) dendogram of the community merging; d) time needed for each pair of oscillators to synchronize. Red for shorter times, blue for larger times.}
\label{RB}
\end{figure}

In a different scenario, there is a set of deterministic networks
that has been used as an example of hierarchical scale-free
networks, proposed by Ravasz and Barabasi \cite{RB}. This type of
networks, apart from its hierarchical structure has some nodes
with a special role in terms of number of connexions (hubs) in
contrast to the networks discussed previously that are essentially
homogeneous in degree. 
In Fig. \ref{RB}a we present a very simple example of this class of networks for the
case of two hierarchical levels. 


In a previous work \cite{nos} we represented the correlation matrix of the system $\rho_{ij} (t)$ at the same time instant $t$ for two slightly different two level hierarchical networks with structure 13-4 and 15-2. From that representation, we could identify the two levels of the hierarchical distribution of communities. The network 13-4 is very close to a state in which the four large groups are almost synchronized whereas the network 15-2 still presents some of the smaller groups of
synchronized oscillators, and the larger group starting to synchronize, coherently with their topological structure. This picture that relates dynamics and topology, and distinguishes at a given time the two configurations, was our starting point and we follow
this formalism in the next section.

\section{The connection between synchronization and topology}

The visualization of the correlation matrix of the system
helps in elucidating the topology of the network. To extract the quantitative information it is useful
to introduce some threshold $T$ to convert the correlation matrix into a binary matrix, that will be used to determine the borders between different groups. We define a {\em dynamic  connectivity}
matrix
 \begin{equation}
{\mathcal D}_t (T)_{ij}=
\left\{
\begin{array}{ll}
1 & \mbox{if } \rho_{ij}(t)>T \\
0 & \mbox{if } \rho_{ij}(t)<T
\end{array}
\right.
 \label{ro_threshold}
\end{equation}
that depends on both the underlying topology and the collective
dynamics. For a fixed time $t$, by moving the threshold $T$, we
obtain different representations of ${\mathcal D}_t(T)$ that inform
about the structure of the dynamic correlations. When the
threshold is large enough the representation of ${\mathcal D}_t(T)$
becomes a set of disconnected clumps or communities. Decreasing
$T$ a hierarchical structure of communities is devised. Note that
since the function $\rho_{ij}(t)$ is continuous and monotonic
(because the existence of a unique attractor of the dynamics), we
can redefine ${\mathcal D}_T(t)$, i.e. fixing the threshold and
evolving in time. We obtain the same information about the
structure of the dynamic connectivity matrix at different time
scales. These time scales unravel the topological
structure of the connectivity matrix at different topological
scales \cite{nos}.

For instance, the eigenvalue spectrum of ${\mathcal D}_T(t)$, $S({\mathcal
D}_T(t))$, can help in tracing the hierarchy of communities. In particular, the number of
null eigenvalues corresponds to the number of connected components of the dynamical
(synchronized) network. Thus, 
at short times, all units are uncorrelated and then we have $N$
disconnected sets, being $N$ the number of nodes in the network.
As time goes on, nodes become synchronized in groups according to
their topological structure. 
In panel b) of Figs. 1-3 we have plotted the number of non-connected components as a function of time for the three networks introduced in the previous section. 
At early stages the number of components is equal to the number of nodes whereas at final stages we have a single component. One of the remarkable results is the slope in these curves. Plateau regions indicate the relative stability of the dynamics at a given time scale whereas a vertical drop is related to instability. In \cite{nos} we already linked the stabililty of these regions with the eigenvalue spectrum of the Laplacian matrix, that we will investigate in detail later on.

Another interesting link between dynamics and topology, 
related with the discussion in the previous paragraph,  
comes from the analysis of the whole spectrum of the
Laplacian matrix of the network graph ${\mathcal L}$ \cite{biggs}. The
Laplacian matrix is defined as ${\mathcal
L}_{ij}=k_i\delta_{ij}-a_{ij}$, where $k_i$ is the degree of node
$i$, $\delta_{ij}$ is the Kronecker delta and $a_{ij}$ is the
element of the adjacency matrix (1 if nodes $i$ and $j$ are
connected and 0 otherwise). The spectral information of the
Laplacian matrix has been used to understand the structure of
complex networks \cite{barabasispect}, and in particular to detect
the community structure \cite{donetti,capocci} (also the spectral analysis of the modularity matrix \cite{PNASNewman} can be used to this end). Recent studies
have also focused on the spectral information of the Laplacian
matrix and the synchronization dynamics
\cite{barahona,motter1,yamir,hong,motter2,lee,munozprl,chavez}.
The common approach is to take advantage of the master stability
equation \cite{pecora} to determine the relation between the
relative stability of the synchronized state (via the ratio
$\lambda_N/\lambda_2$) and the heterogeneity of the topology,
although sometimes some language abuse appears and authors talk
about better or worse synchonizability instead of stability of the
synchronized state. Our approach differs from these works in the
following: we are interested in the transient towards
synchronization because it is this whole process which will reveal
the topological structure at different scales. For this reason our
analysis focus on the whole eigenvalue spectrum of the Laplacian
matrix $S({\mathcal L})$.
To characterize this spectrum, we rank the eigenvalues of ${\mathcal
L}$ using an index $i$ in ascending order $0=\lambda_1\le
\lambda_2\le \ldots \lambda_i \ldots \le \lambda_N$. 
Panel b) of Figs. 1-3 shows the order of the eigenvalues in this ranking
versus the inverse of its values, which closely corresponds to the time scales of emergence of nested communities of nodes. The structure
of this sequence brings to light many aspects of the topological
structure:  (i) the number of null eigenvalues gives trivially the
number of disconnected components of the static network, (ii) the gaps between
consecutive eigenvalues tell us about the relative differences of
time scales, and (iii) large eigenvalues in the last part of the
series stands for the existence of hubs in the network (we will
turn to these points later).  In any case case the two curves on each of the panels b)
show an intriguing similarity. The time scale related to a given eigenvalue corresponds roughly to the 
time scale at which a number of groups of oscillators are synchronized 
(which is shown by the number of connected components of the dynamical matrix).
%

In panel c) of Figs. 1-3 we have plotted the dendogram of the merging of the groups along the time evolution as it can be deduced from the time evolution of the correlation matrix. Groups merge as they get more correlated above some threshold (synchronized). We can clearly identify here the different topological scales, i.e. communities at different hierarchical levels.

Finally, in panel d) we plot the time that each pair of nodes need to synchronize. This picture complements the previous one, since it did not preserve  the transitivity character of the merging. In the current picture transitivity plays no role and we can observe how pairs of nodes synchronize by itself and not by means of third parties.

\section{Linear analysis}

Finally we would like to shed some light about the intriguing
relationship between the  eigenvalues of the Laplacian and the
dynamic structures that emerge in the route towards synchronization as shown in panel b) of Figs. 1-3.
To understand this correspondence let us analyze the linearized
dynamics of the Kuramoto model (i.e. the dynamics close to the
attractor of synchronization) in terms of the Laplacian matrix,
\begin{equation}
\frac{d\theta_i}{dt}=-k\sum_{j}
L_{ij} \theta_j \hspace{0.5cm} i=1,...,N
 \label{linearmodel}
\end{equation}
whose solution in terms of the normal modes $\varphi_i(t)$ reads
\begin{equation}
\varphi_i(t)=\sum_{j} B_{ij}\theta_j=
\varphi_i(0) e^{-\lambda_i t}\hspace{0.5cm} i=1,...,N
 \label{linearsolution}
\end{equation}
where $\lambda_i$ are the eigenvalues of the
Laplacian matrix, and $B$ is the matrix of eigenvectors.
This set of equations has to be satisfied at any time $t$. If we
rank the system of equations in descending order of the
eigenvalues (i.e. starting from $\lambda_N$), the right hand side
system of Eq.(\ref{linearsolution}) will approach zero in a
hierarchical way. This fact is equivalent in the dynamics to group
oscillators surpassing the synchronization threshold forming
communities. The gaps in the spectrum $S({\mathcal L})$ represent
clearly different time scales between modes revealing different
topological scales. The collective modes, solution of the system
represented by Eq.(\ref{linearsolution}), denote two types of
behaviors. Some modes provide information about reorganization of
the phases in the whole network, while the others inform about
synchronization between pairs or groups of oscillators. The
presence of hubs in the topology gives rise to large eigenvalues
that decay very fast and are related to the first type of modes,
those representing "synchronization" between the hub and the {\em
topological average} of the phases of rest of oscillators. The
rest of modes relate oscillators that have similar projections on
the corresponding eigenvectors thus giving rise to communities at
a given topological scale. Indeed, this fact supports the success
of the identification of communities using spectral analysis
\cite{donetti}.

To illustrate these ideas we have explicitly analyzed two different
types of networks: the star, which is a simple
example of isotropic, homogeneous network without inner structure,
and a simplified version of the Ravasz-Barabasi network, where we
can easily identify all the time scales of the dynamic process and
also to characterize and interpret the information provided by the
spectrum of eigenvalues and eigenvectors of the laplacian matrix.
The star consists of a network formed by two types of nodes: a hub,
located in the middle of the network, connected to the rest of
nodes and the peripherical nodes, not connected between them, only
to the hub. It has been studied frequently in communication
problems as a paradigm of system where all the traffic goes
through a single node and therefore it is easy to collapse. The
laplacian matrix of a n-node star (one hub and n-1 peripherical
nodes) is
\begin{equation}
\left(
  \begin{array}{ccccc}
    -(n-1) & 1 & \cdots & 1 & 1 \\
    1 & -1 & \cdots & 0 & 0 \\
    \vdots & \vdots & \vdots & \vdots & \vdots \\
    1 & 0 & \cdots & -1 & 0 \\
    1 & 0 & \cdots & 0 & -1 \\
  \end{array}
\right)
\end{equation}
whose spectrum of eigenvectors has a very simple and compact
structure:
\begin{equation}
(0,-1,\ldots,-1,-n)
\end{equation}
\noindent two non-degenerate eigenvalues $(0, -n)$ and a $(n-2)$
times degenerate eigenvalue $(-1)$. The largest eigenvalue, in
absolute value, gives information about the dynamical properties
of the hub, while the smallest characterizes the attractor. To
understand how the system of oscillators approaches the
synchronized final state it is convenient to analyze the spectrum
of eigenvectors in a hierarchical way starting again from the mode
decaying first

\begin{equation}
\begin{array}{ccccc}
       \lambda=-n &\hspace{1cm} &v_1&=&(-(n-1),1,1,1,\ldots,1,1,1)\\
       \lambda=-1 &\hspace{1cm} &v_2&=&(0,-1,1,0,\ldots,0,0)\\
       \lambda=-1 &\hspace{1cm} &v_3&=&(0,-1,0,1,\ldots,0,0) \\
       \vdots &\hspace{1cm}& \ldots \\
       \lambda=-1 &\hspace{1cm} &v_{(n-1)}&=&(0,-1,0,\ldots,0,0,1) \\
       \lambda=0 &\hspace{1cm} &v_n&=&(1,1,1,1\ldots,1,1,1)
     \end{array}
\end{equation}
The dynamical evolution of the system can be interpreted as
follows. The equation related to the largest eigenvalue reads
\begin{equation}
\sum_j B_{1j}\theta_j=e^{-nt}
\end{equation}
It means that in a first stage, almost instantaneously, the phase of
the hub goes to
\begin{equation}
\theta_1= \frac{1}{n-1} \sum_{i>1} \theta_i
\end{equation}
\noindent that is an arithmetic average over the peripherical units.
In a second stage, the rest of phases synchronize in pairs $\theta_i=\theta_j \hspace{1ex} 
\forall i,j \ne 1$.
Therefore, we can identify two different types of collective modes,
one tending to reorganize the phases of some units while the rest
describe the physical entrainment between oscillators. Notice that
the synchronization process takes place in a single time scale which
is reflected in the spectral properties of the laplacian matrix as
the set of eigenvalues is multiply degenerate.

A more complex situation is presented in the simplified
Ravasz-Barabasi network Fig.\ref{RB}a. In contrast to the
uniform structure of the star, this network displays an internal
topological structure in two levels. Therefore, we expect a rich
dynamical evolution which also should be reflected in the spectrum
of eigenvalues of the laplacian matrix. Now, the laplacian matrix
reads
\begin{equation}
\left(
  \begin{array}{ccccccccc}
    -8 & 1 & 1 & 1 & 1 & 1 & 1 & 1 & 1 \\
    1 & -3 & 0 & 0 & 0 & 1 & 1 & 0 & 0 \\
    1 & 0 & -3 & 0 & 0 & 0 & 0 & 1 & 1 \\
    1 & 0 & 0 & -2 & 1 & 0 & 0 & 0 & 0 \\
    1 & 0 & 0 & 1 & -2 & 0 & 0 & 0 & 0 \\
    1 & 1 & 0 & 0 & 0 & -3 & 1 & 0 & 0 \\
    1 & 1 & 0 & 0 & 0 & 1 & -3 & 0 & 0 \\
    1 & 0 & 1 & 0 & 0 & 0 & 0 & -3 & 1 \\
    1 & 0 & 1 & 0 & 0 & 0 & 0 & 1 & -3 \\
  \end{array}
\right)
\end{equation}
\noindent whose spectrum of eigenvalues and eigenvectors is
\begin{equation}
\begin{array}{ccccc}
     \lambda=-9 &\hspace{0.5cm}& v_1&=&(-8, 1, 1, 1, 1, 1, 1, 1, 1)\\
     \lambda=-4 &\hspace{0.5cm}&v_2&=&(0,0,-1,0,0,0,0,1,0)\\
     \lambda=-4 &\hspace{0.5cm}&v_3&=&(0,0,-1,0,0,0,0,0,1)\\
     \lambda=-4 &\hspace{0.5cm}&v_4&=&(0,-1,0,0,0,1,0,0,0)\\
     \lambda=-4 &\hspace{0.5cm}&v_5&=&(0,-1,0,0,0,0,1,0,0)\\
     \lambda=-3 &\hspace{0.5cm}&v_6&=&(0,0,0,-1,1,0,0,0,0) \\
     \lambda=-1 &\hspace{0.5cm}&v_7&=&(0,1,0,\frac{-3}{2},\frac{-3}{2},1,1,0,0) \\
     \lambda=-1 &\hspace{0.5cm}&v_8&=&(0,0,1,\frac{-3}{2},\frac{-3}{2},0,0,1,1) \\
     \lambda=0 &\hspace{0.5cm}&v_9&=&(1,1,1,1,1,1,1,1,1)
     \end{array}
\end{equation}
The analysis of the dynamical relaxation of the eigenmodes shows the
following aspects. First, a reorganization process takes place
involving the phases of the hub with the rest of oscillators. This
stage is similar to that observed in the star because the hub is
again is a fully connected unit. The four-fold degenerate eigenvalue
$\lambda=-4$ describes the entrainment process between 6 nodes. This
behavior is quite appealing because although they are not directly
connected they are topologically equivalent and belong to the same
structural level. These results show that one has to consider other
elements more important than physical connectivity when defining the
concept of community in complex networks. Later, at a slightly
larger time scale units 4 and 5 synchronize. Finally these nodes
synchronize with the two equivalent functional groups as can be
stated from the two modes associated to the eigenvalue $\lambda=-1$.
The results are in perfect agreement with the synchronization times
of the full non-linear dynamics as we can see in the dendrogram of
Fig. 3.

In general, as a rule of thumb, those networks whose structure is
random and where all the nodes play the same statistical role from a
topological standpoint are characterized by a continuum spectrum of
eigenvalues. In contrast, when the network has a specific internal
structure, which for social or biological networks might be the
fingerprint of different functional groups, then we expect a
Laplacian spectrum with different gaps stressing the existence of
different time scales in the dynamic process towards the attractor.

\section{Summary}

We have analyzed the synchronization dynamics in
complex networks and show how this process unravels its different
topological scales. We have also reported a connection between the
spectral information of the Laplacian matrix and the hierarchical
process of emergence of communities at different time scales.
We show for a set of structured networks that the gaps in the spectrum are
related to the stability of the hierarchical structures (communities) of the networks.
We propose additional graphical tools to understand how the synchronization process
takes place: a dendogram of the merging of the synchronized groups and a visualtization of the
time needed for each pair of oscillators to synchronize.

\section*{Acknowledgments}
This work has been supported by DGES of the Spanish Government Grant No. BFM-2003-08258.


\begin{thebibliography}{00}
\bibitem{footnote}
We prescribe the value of the threshold $T$ to 0.999. Other choices do not alter the results, since it only modifies the relative time scales.
\end{thebibliography}
\end{document}